\input{aipcheck}

\documentclass[
    ,final            
  ]
  {aipproc}

\layoutstyle{6x9}

\begin{document}

\title{Experimental Highlights of the RHIC Program}


\author{Patricia Fachini}{
  address={Brookhaven National Laboratory, P.O. BOX 5000, Upton, NY, USA.}
}

\begin{abstract}
Experimental highlights of the RHIC program are reviewed.
\end{abstract}

\maketitle


\section{Introduction}

Relativistic heavy-ion collisions provide a unique environment to
study matter under extreme conditions of high temperature and
energy density. RHIC (Relativistic Heavy Ion Collider) at the
Brookhaven National Laboratory in Upton, NY provides us with such
collisions. Since the beginning of the RHIC operation in the year
2000, RHIC has provided us not only with Au+Au collisions at the
top energy of $\sqrt{s_{_{NN}}}$ = 200 GeV, but also at 62 GeV.
RHIC has also provided collisions of lighter systems such as Cu+Cu
at $\sqrt{s_{_{NN}}}$ = 200, 62, and 22.5 GeV, and finally the
$p+p$ and $d$+Au collisions at $\sqrt{s_{_{NN}}}$ = 200 GeV that
were used as references.

The amount of data obtained so far is as overwhelming as the
results. I will try to summarize where we are, with RHIC RUNV just
about to start.

\section{Elliptic Flow}

In non-central collisions the initial spatial anisotropy is
transformed into an anisotropy in momentum-space if sufficient
interactions occur among the constituents within the system. Once
the system has expanded enough to quench the spatial anisotropy,
further development of momentum anisotropy ceases. This
self-quenching process happens quickly, so elliptic flow is
primarily sensitive to the early stages of the collisions
\cite{huo}.

\subsection{Hydrodynamics}

\begin{figure}
  \includegraphics[height=.3\textheight]{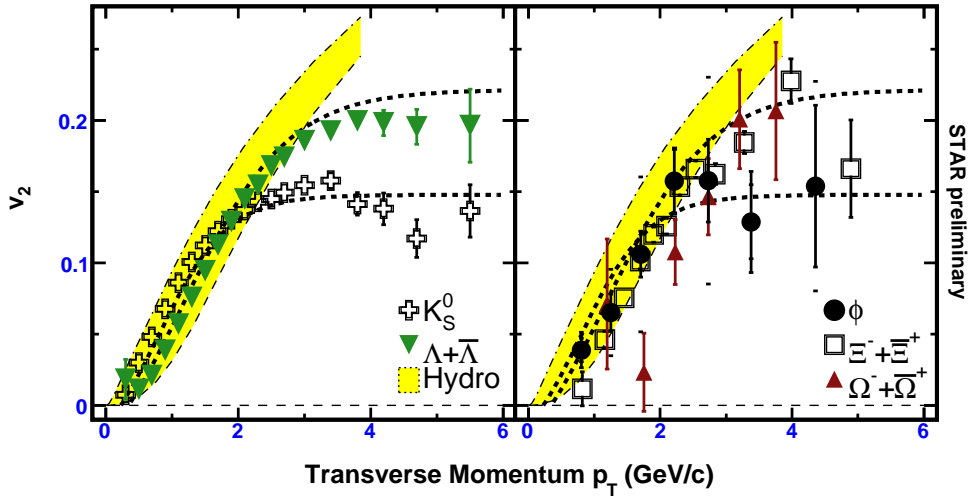}
  \caption{Azimuthal anisotropy $v_2$ for strange (left panel) and
  multi-strange (right panel) hadrons in minimum bias Au+Au collisions
  \cite{old}. Data measured by STAR.
  The dashed lines show a common fit to the $K^0_S$ and $\Lambda +
  \overline{\Lambda}$ data \cite{dong}. The shaded areas are hydrodynamic
  calculations \cite{huo}}\label{fig:flow2}
\end{figure}

The elliptic flow $v_2$ as a function of $p_t$ for $K^0_S,
\Lambda, \phi, \Xi$, and $\Omega$ is depicted in Fig.
\ref{fig:flow2} \cite{old,cai}. The $\phi, \Xi$, and $\Omega$ have
low hadronic cross-sections, therefore the large $v_2$ observed
suggest that the elliptic flow is built up in the partonic stage.
The expected range of $v_2$ from hydrodynamic calculations is also
shown in Fig. \ref{fig:flow2}. A more detailed comparison can be
seen in Fig. \ref{fig:flow1} (left panel), where  the mass
dependence hydrodynamic results \cite{huo} are compared to the
$v_2$ measurements of $\pi, K, p$, and $\Lambda$ \cite{prc72}.
Hydrodynamics describes well the mass dependence observed in the
data that is characteristic of a common flow velocity. Since an
ideal hydrodynamic fluid is a thermalized system with a zero mean
free path that yields to the maximum possible $v_2$, the good
agreement between the measured $v_2$ and the hydro results
\cite{huo} suggests thermalization in heavy-ion collisions at
RHIC.

\subsection{Constituent Quark Scaling}

\begin{figure}
  \includegraphics[height=.29\textheight]{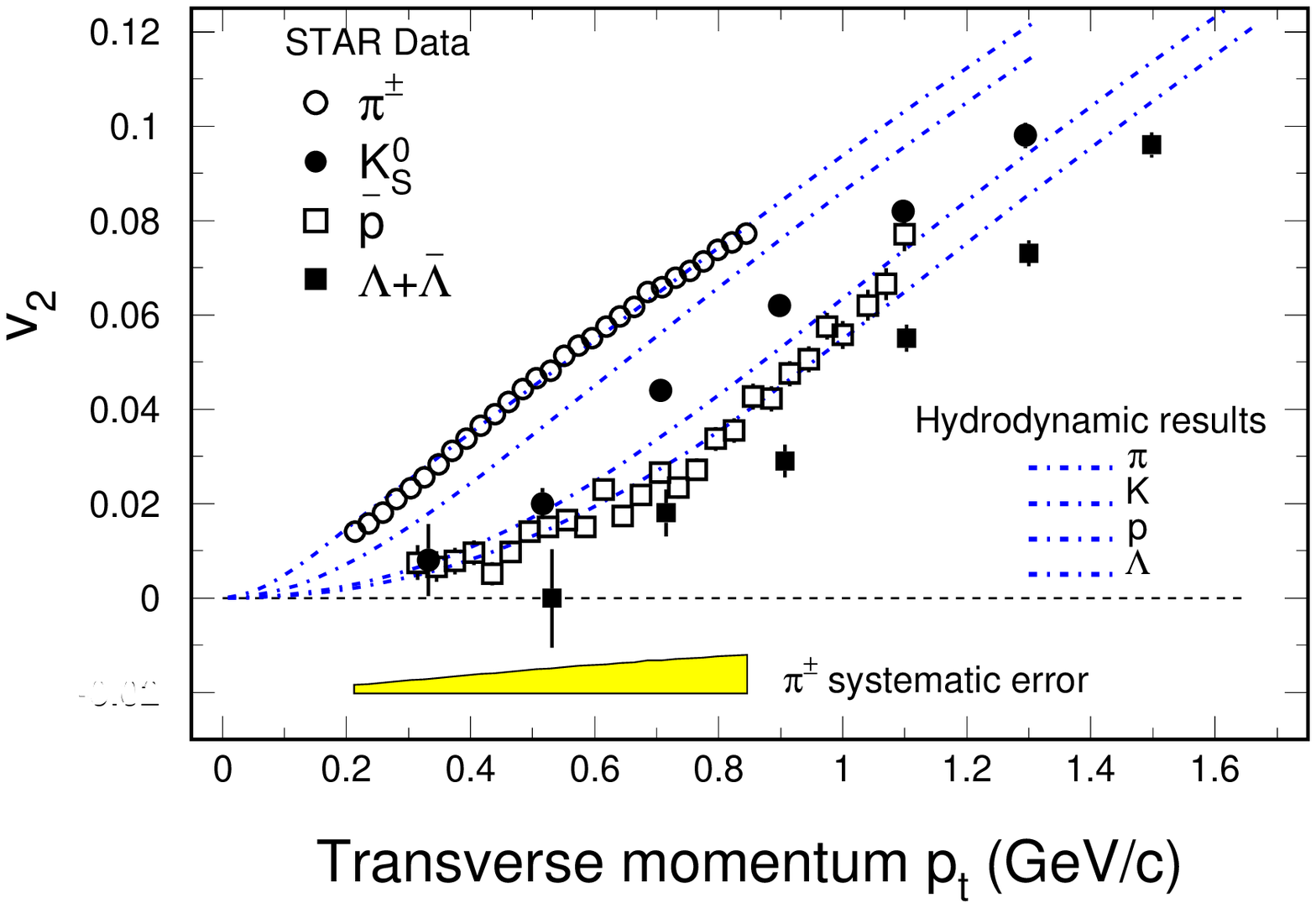}
  \includegraphics[height=.35\textheight]{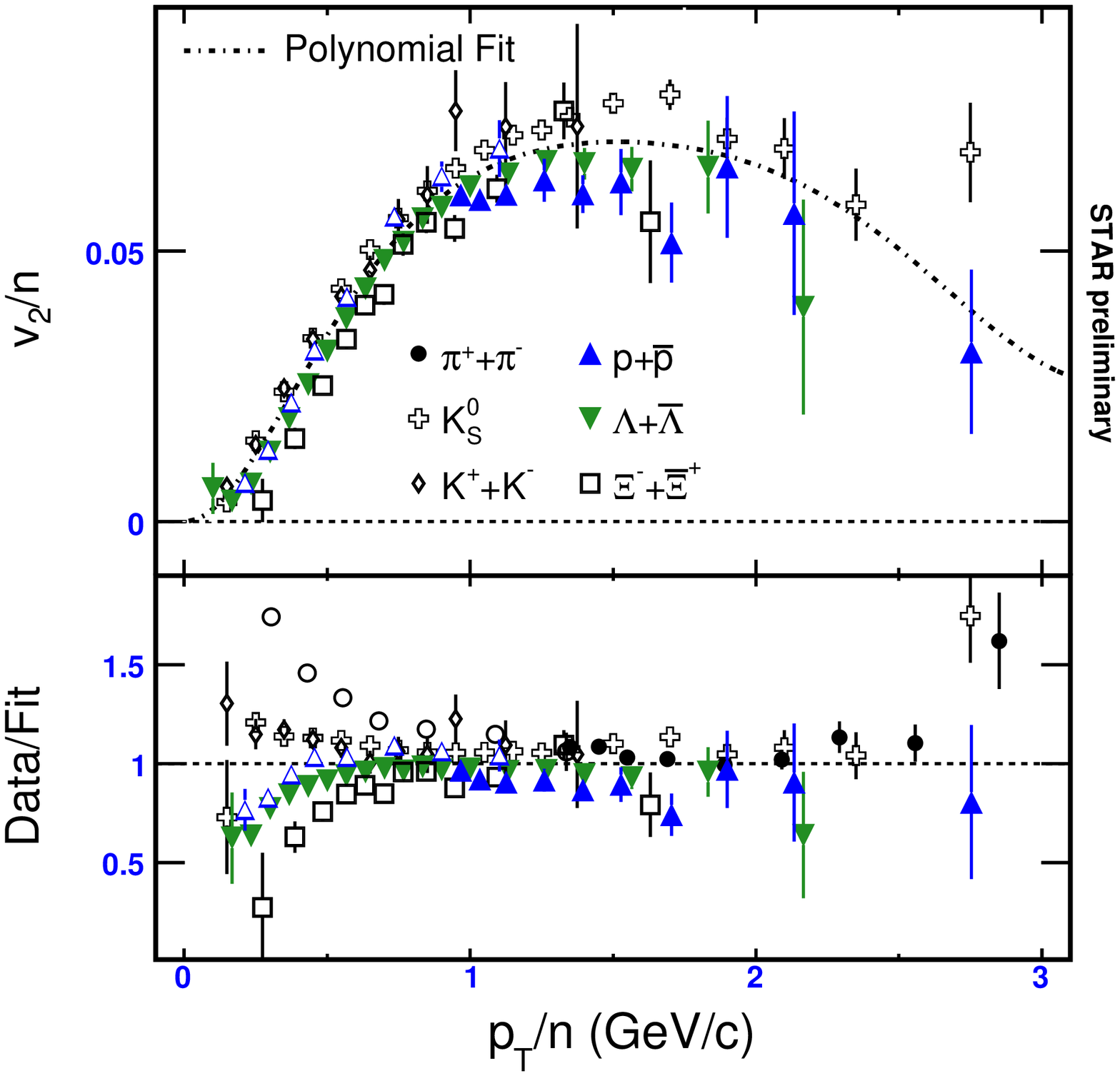}
  \caption{Left panel: $v_2$ as a function of $p_t$ from minimum
  bias Au+Au collisions \cite{prc72} measured by STAR. The dotted-dashed lines are
  hydrodynamic calculations using an equation of estate (EOS) with a first-order
  hadron quark-gluon plasma phase transition \cite{huo}. The description of the
  data worsen if a hadronic EOS is used \cite{huo}.  Right panel:
  Measurements by STAR
  of the scaled $v_2(p_T/n)/n$ for identified hadrons (upper panel) and the ratio between
  the measurements and a polynomial fit through all data points (lower panel) except
  the pions for $\sqrt{s_{_{NN}}}$ = 200 GeV minimum bias Au+Au collisions \cite{old}.}
  \label{fig:flow1}
\end{figure}

While hydrodynamic calculations keep increasing as a function of
$p_T$, the measured $v_2$ saturate at $p_T > $ 2 GeV/$c$
\cite{old}. The saturation value for mesons is about 2/3 of that
for baryons. This separation pattern holds for $\pi, K, \Lambda$,
and $\Xi$, and seems to hold for $\phi$ and $\Omega$
\cite{cai,prl92}. This result and the baryon-meson splitting of
the high $p_T$ suppression pattern \cite{dun} suggest the
relevance of the constituent quark degrees of freedom in the
intermediate $p_T$ region \cite{mol}. $v_2$ scaled by the number
of valence quarks $n$ as a function of $p_T/n$ is depicted in Fig.
\ref{fig:flow1} (right panel). The lower panel of Fig.
\ref{fig:flow1} (right panel) displays the ratio between the
measurements and a polynomial fit to all the data. At low $p_T/n$
the observed deviations from the fit follow a mass ordering which
is expected from hydrodynamics. At higher $p_T$, all $v_2/n$
measurements are reasonably close to unity showing the constituent
quark scaling.

\section{Freeze-Out Properties}

The measured particle spectra and yields \cite{prl921} and
event-by-event $\langle p_T \rangle$ fluctuations \cite{ada}
indicate a nearly chemically and kinetically equilibrated system
at the final freeze-out stage.

\subsection{Chemical and Kinetic Freeze-out Parameters}

\begin{figure}
  \includegraphics[height=.26\textheight]{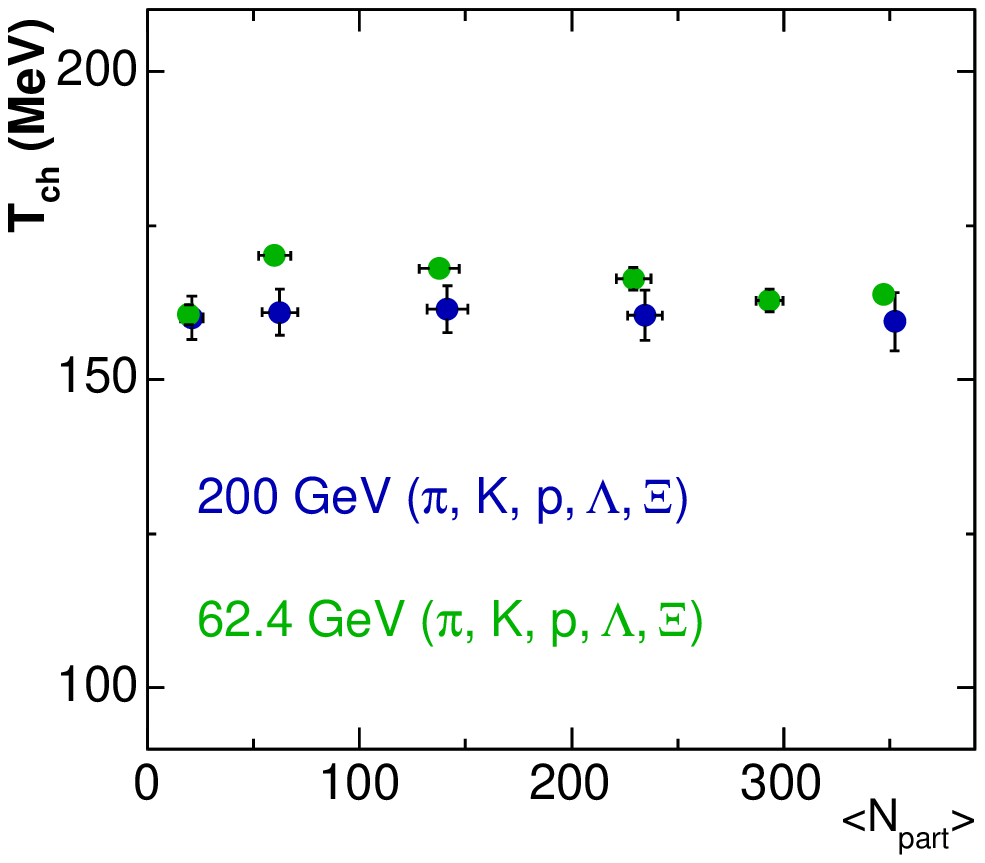}
  \includegraphics[height=.26\textheight]{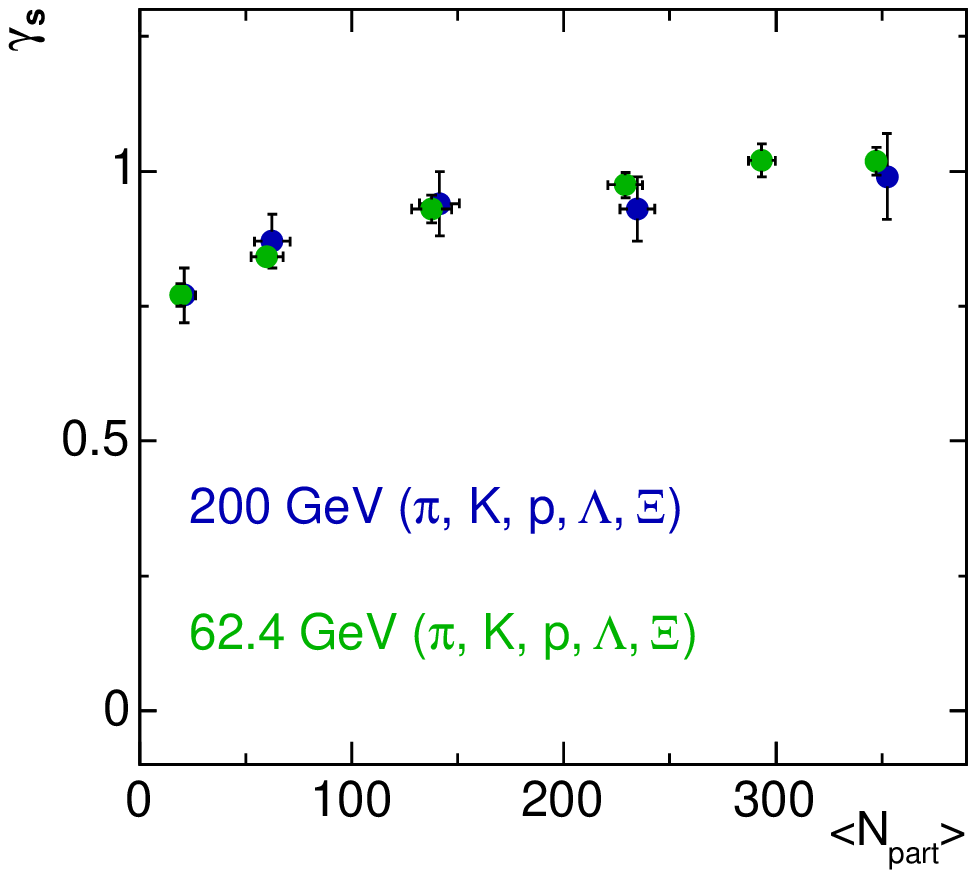}
  \caption{Left panel: Extracted chemical freeze-out temperature from stable particle
  ratios \cite{mol2,spe}.Right panel: Extracted strangeness suppression factor from stable
  particle ratios \cite{mol2,spe}. STAR measurements.}\label{fig:temperature}
\end{figure}

\begin{figure}
  \includegraphics[height=.35\textheight]{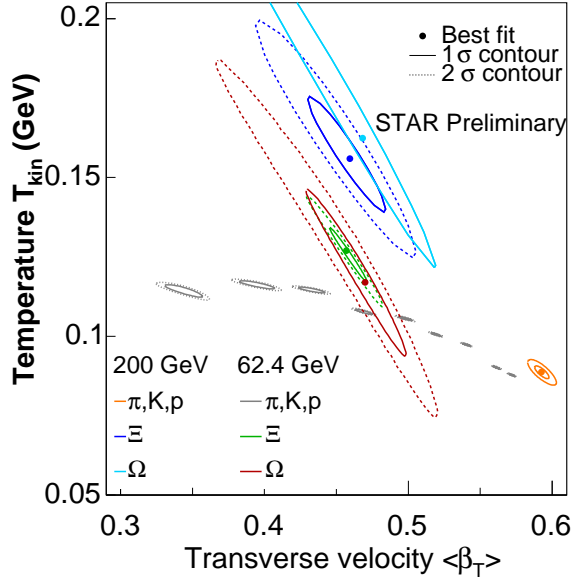}
  \caption{Extracted kinetic freeze-out temperature as a function of the average
  flow velocity within the blast-wave model \cite{mol2,spe}.
  STAR measurements.}\label{fig:blastwave}
\end{figure}

STAR has now measured hadron distributions at $\sqrt{s_{_{NN}}}$ =
200  and 62 GeV \cite{mol2,spe}. Chemical freeze-out properties
were extracted from stable particle ratios within the thermal
model \cite{pbm}.  The extracted chemical freeze-out temperature
$T_{ch}$ and the strangeness suppression factor $\gamma_s$ are
depicted in Fig. \ref{fig:temperature}. Kinetic freeze-out
properties from particle $p_T$ distributions were extracted within
the blast wave model \cite{sch}. Figure \ref{fig:blastwave}
displays the extracted kinetic freeze-out temperature $T_{kin}$
and the average radial flow velocity $\langle \beta_T \rangle$.
The results at $\sqrt{s_{_{NN}}}$ = 62 GeV are found to be
qualitatively the same as those obtained at $\sqrt{s_{_{NN}}}$ =
200 GeV and resonance decays are found to have no significant
effect on the extract kinetic freeze-out parameters \cite{spe}.

$T_{ch}$ is independent of centrality. $T_{kin}$ obtained from
$\pi, K,$ and $p$ decreases as a function of centrality, while the
corresponding $\langle \beta_T \rangle$ increases. This is
evidence that the system expands between chemical and kinetic
freeze-outs, which brings the system to a lower temperature. The
constant $T_{ch}$ suggests that hadronic scatterings from
hadronization to chemical freeze-out may be negligible because
they would result in a dropping $T_{ch}$.

$\gamma_s$ increases from $p+p$ to peripheral and central Au+Au
collisions. In central Au+Au collisions, $\gamma_s$ is $\sim$1
suggesting that strangeness is saturated. The $T_{ch}$ of $\phi,
\Xi,$ and $\Omega$ is higher than that of $\pi, K,$ and $p$, while
the $\langle \beta_T \rangle$ is lower. Noting that the $\phi,
\Xi,$ and $\Omega$ have small hadronic cross-sections, they may
chemically and kinetically freeze-out at the same time.

\section{System Size Dependence}

\begin{figure}
  \includegraphics[height=.32\textheight]{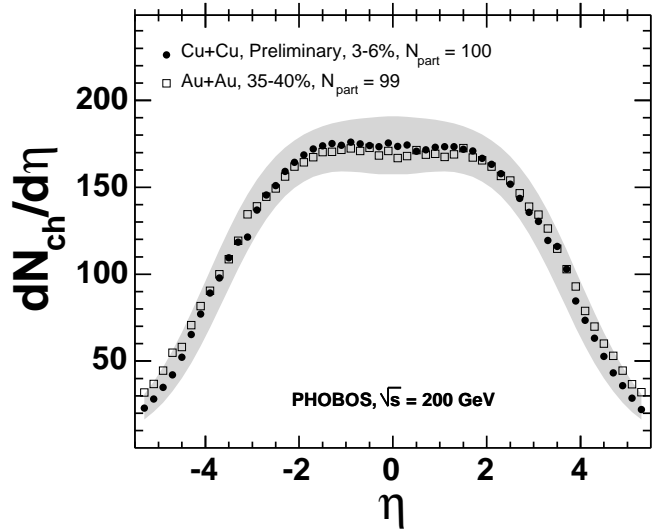}
  \caption{Charged hadron pseudorapidity measured by PHOBOS
  for Cu+Cu and Au+Au
  collisions at $\sqrt{s_{_{NN}}}$ = 200 GeV with similar number
  of participants $N_{part}$
  }\label{fig:phobos}
\end{figure}

The comparison between the charged hadron pseudorapidity
distributions measured by PHOBOS \cite{rol} in Cu+Cu and Au+Au
with similar number of participants $N_{part}$ is presented in
Fig. \ref{fig:phobos}. Both distributions are comparable within
errors, showing that bulk particle production depends mainly on
the number of participating nucleons. The same is true for
different $N_{part}$ and also at $\sqrt{s_{_{NN}}}$ = 62 GeV
\cite{rol}. Many other measurements from BRAHMS, PHENIX, PHOBOS,
and STAR support this argument \cite{sta,gre,rol,dun}.

\section{Low $p_T$ Hadrons}

\begin{figure}
  \includegraphics[height=.26\textheight]{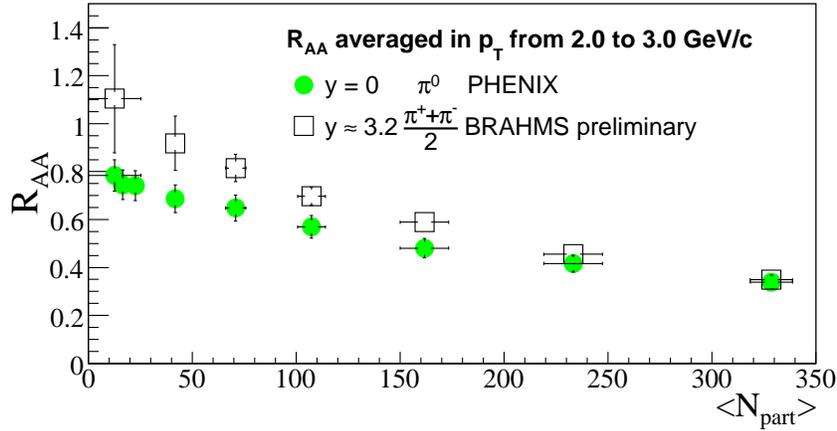}
  \caption{Averaged $R_{AA}$ in the interval 2.0 $ < p_T < $ 3.0 GeV/$c$ at mid-rapidity
  (PHENIX) and at forward rapidity (BRAHMS) as a function of number of participants at
  $\sqrt{s_{_{NN}}}$ = 200 GeV \cite{sta}}.\label{fig:raanpart}
\end{figure}

The nuclear modification factor $R_{AA}$ of $\pi$ and $p$ measured
at $y =$ 3.2 by BRAHMS and $y =$ 0 by PHENIX in central Au+Au
collisions does not depend on rapidity \cite{sta}, suggesting that
the same mechanisms is responsible for the nuclear modifications.
It has been predicted \cite{gyu} that the magnitude of jet
quenching should depend on both the size and the density of the
created absorbing medium. The averaged pion $R_{AA}$ as a function
of $N_{part}$ for both forward rapidity and mid-rapidity are shown
in Fig. \ref{fig:raanpart}. The average was performed in the
interval 2 $ < p_T < $ 3 GeV/$c$. The mid- and forward rapidity
pion suppression for the most central Au+Au collisions are found
to be the same magnitude. However, the $R_{AA}$ measured in
forward rapidity shows significantly stronger rise towards
peripheral collisions as compared to $R_{AA}$ at mi-rapidity,
differing on the level of 35$\%$ for $\langle N_{part} \rangle
\approx$ 100. This is consistent with the model of parton energy
loss in a strongly absorbing medium \cite{dai,dre}. In this
picture, at mid-rapidity, the emission is dominated by the
emission from the surface which quenches the dependence of
$R_{AA}$ on the system. On the other hand, at forward rapidities,
the transition from surface to volume emission can occur, which
leads to a stronger dependence on the number of participants.

\section{Intermediate $p_T$ Hadrons}

\begin{figure}
  \includegraphics[height=.27\textheight]{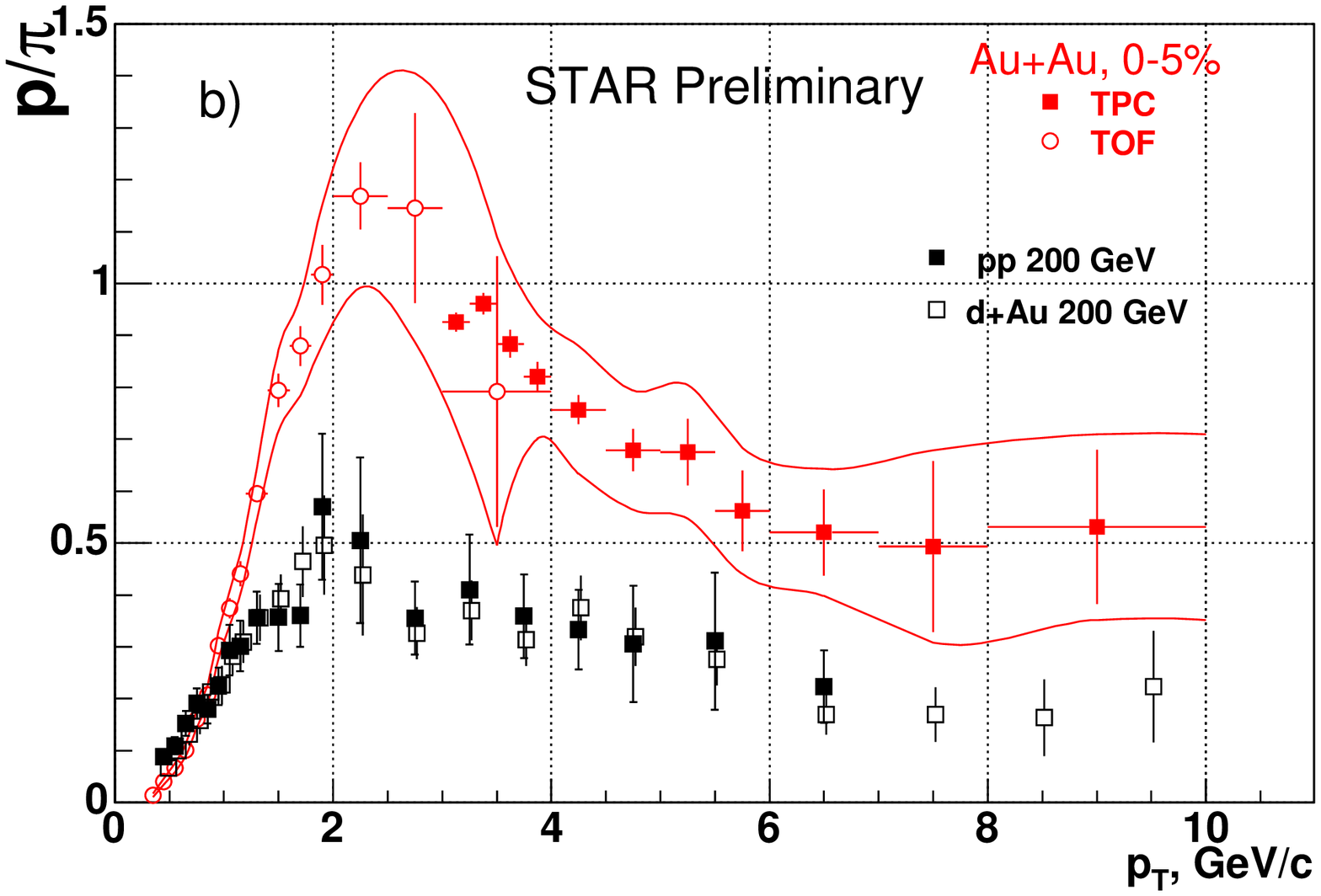}
  \includegraphics[height=.271\textheight]{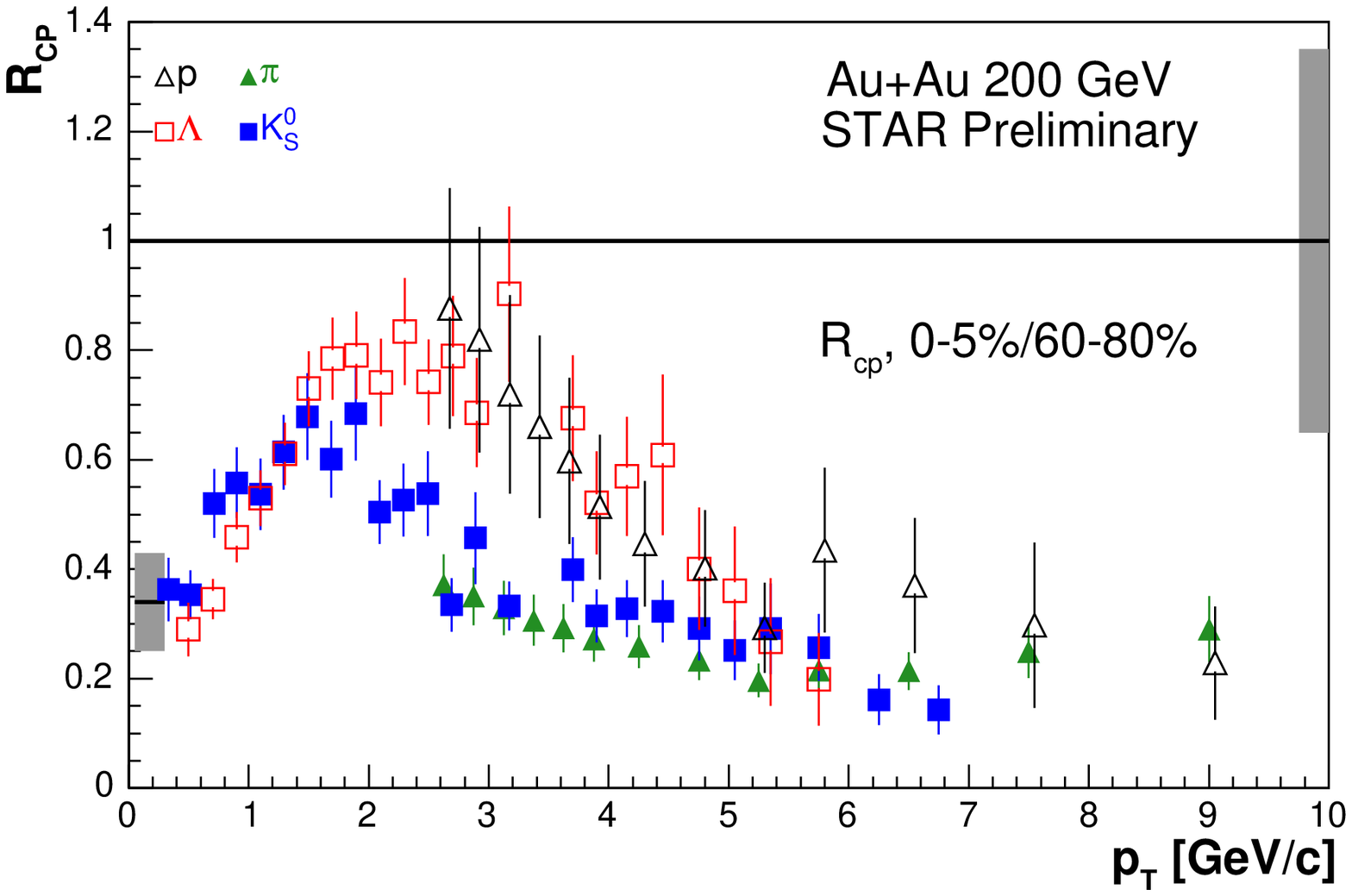}
  \caption{Left panel:$p/\pi$ ratio measured by STAR as a function of $p_T$ for central Au+Au
  (0-5$\%$), $d$+Au and $p+p$ collisions \cite{bar}. For the Au+Au measurements, the
  error bars are statistical only and the solid lines are the systematic
  uncertainties. For the $d$+Au and $p+p$ measurements, the errors shown
  are combined statistical and systematic. Right panel: $R_{CP}$ as a function
  of $p_T$ for identified particles measured by STAR \cite{salur,bar}. Errors are statistical
  and systematic. Grey bands are common scale uncertainties from $N_{binary}$.}\label{fig:ppi}
\end{figure}

\begin{figure}
  \includegraphics[height=.35\textheight]{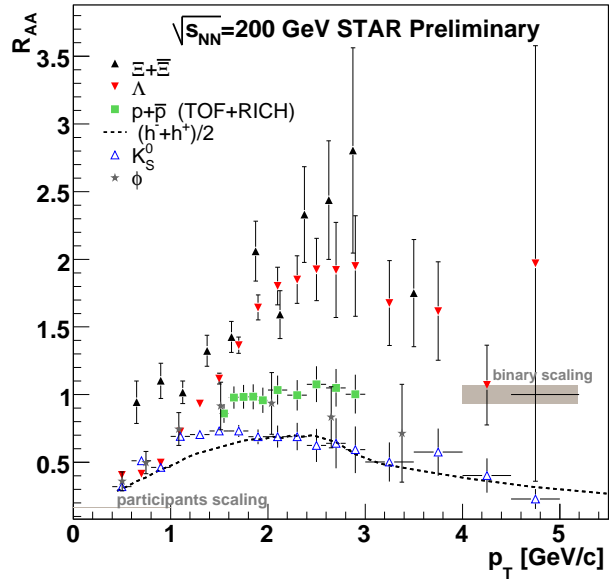}
  \caption{$R_{AA}$ as a function of $p_T$ for identified particles measured by STAR
  \cite{salur,bar}. Errors are statistical and systematic. Grey bands are common
  scale uncertainties from $N_{binary}$.}\label{fig:raas}
\end{figure}

Intermediate $p_T$ ($p_T <$ 6 GeV/$c$) protons behave differently
than mesons at heavy-ion collisions \cite{prl92}. This behavior
can be seen in Fig. \ref{fig:ppi} (left panel) that depicts the
ratio of $p$ to $\pi$ spectra measured by STAR in both central
Au+Au collisions and in $p+p$ collisions. The large enhancement of
the $p/\pi$ ratio at intermediate $p_T$ in Au+Au collisions
indicates that jet fragmentation in vacuum is not the dominant
source of particle production in this $p_T$ range. An enhancement,
roughly  peaked at at the same position,  is also observed in
$R_{CP}$, as shown in Fig. \ref{fig:ppi} (right panel). Here,
$R_{CP}$ is the ratio between central and peripheral Au+Au
collisions normalized to the binary collision scaling expectation.
Like $v_2$ \cite{old}, $R_{CP}$ also separates into baryons and
mesons. The $K^*$ \cite{kstar}, which is a meson with the mass
close to the proton mass, follow the behavior of mesons. This
proves that this separation is not due to the hadron mass. This
grouping is violated in $R_{AA}$, where the reference is from
$p+p$ collisions rather than peripheral collisions. A strong
enhancement in strange baryon $R_{AA}$ with increasing enhancement
for increasing strangeness content is observed at intermediate
$p_T$ \cite{salur}. This enhancement is depicted in Fig.
\ref{fig:raas}.

\section{High $p_T$ Hadrons}

\begin{figure}
  \includegraphics[height=.3\textheight]{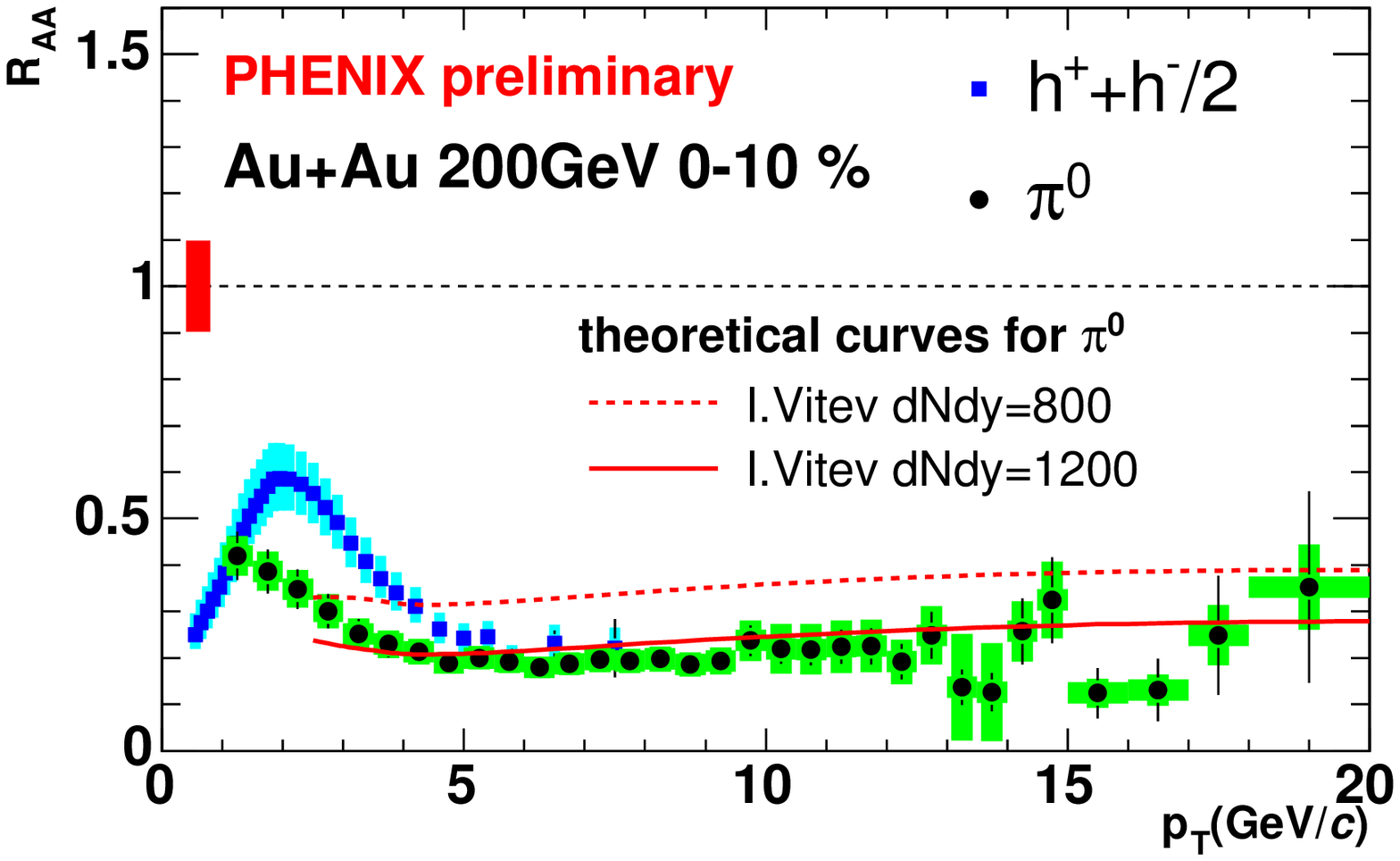}
  \caption{$R_{AA}$ as a function of $p_T$ of $\pi^0$ and charged
  hadrons measured by PHENIX in central Au+Au collisions
  (0-10$\%$) with theoretical predictions \cite{vit,wan}. The shaded areas are
  the systematic uncertainties.
}\label{fig:raapi0}
\end{figure}

\begin{figure}
  \includegraphics[height=.3\textheight]{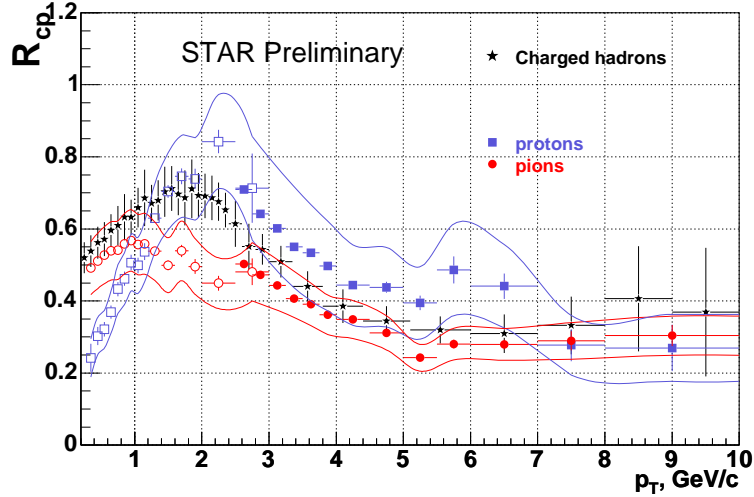}
  \caption{$R_{CP}$ for the 5$\%$ most central collisions measured by STAR,
  normalized by the peripheral 60-80$\%$ collisions, as a function
  of $p_T$ \cite{bar}. The bands show combined statistical and systematic
  errors.
}\label{fig:rcpppi}
\end{figure}

The suppression of high $p_T$ hadrons in central $Au+Au$
collisions was one of the unexpected and important phenomena
observed at RHIC. The $R_{AA}$ of $\pi^0$ measured by PHENIX in
central Au+Au collisions at $\sqrt{s_{_{NN}}}$ = 200 GeV
\cite{shi} is presented in Fig. \ref{fig:raapi0}. The suppression
is quite strong ($R_{AA} \sim$0.2) and remains approximately flat
up to 20 GeV/$c$. Partonic radiative energy loss models
\cite{vit,wan,esk} reproduce this behavior well. In particular,
one calculation depicted in Fig. \ref{fig:raapi0} uses an average
gluon average density $dN_g/dy \sim$ 1200 \cite{vit}. The $R_{AA}$
measured by PHENIX and STAR are shown in Fig. \ref{fig:raapi0}
\cite{shi} and Fig. \ref{fig:rcpppi} \cite{bar}, respectively. The
difference between charged hadrons $R_{AA}$ and the $\pi^0$
$R_{AA}$ in the intermediate $p_T$ region ($p_T <$ 6 GeV/$c$) is
due to the proton contribution as observed in Fig.
\ref{fig:rcpppi} and nicely described by recombination models.
This behavior disappears for $p_T >$ 6 GeV/$c$ and are also
explained by the same recombination models \cite{fri,hwa,gre}.

\section{Jet Correlations}

\begin{figure}
  \includegraphics[height=.3\textheight]{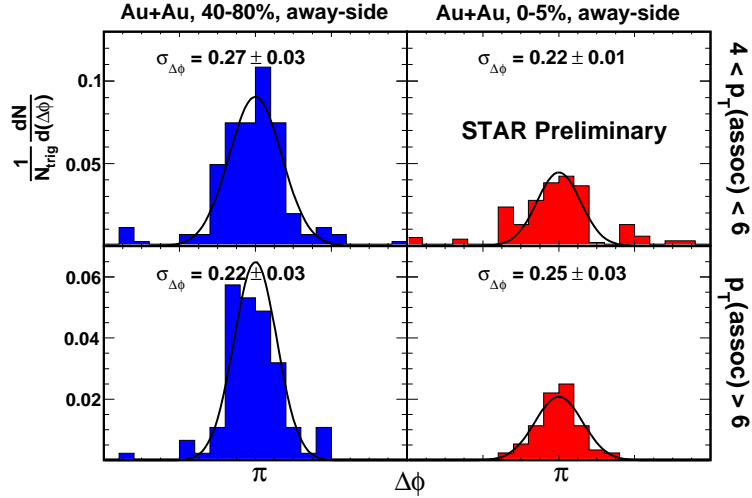}
  \caption{Azimuthal distributions of away-side charged hadrons for
  8 $< p_T^{trig} <$ 15 GeV/$c$ in two different centralities at
  $\sqrt{s_{_{NN}}}$ = 200 GeV Au+Au collisions. These are the total
  distributions, no backgrounds have been subtracted.
  }\label{fig:jet}
\end{figure}

\begin{figure}
  \includegraphics[height=.4\textheight]{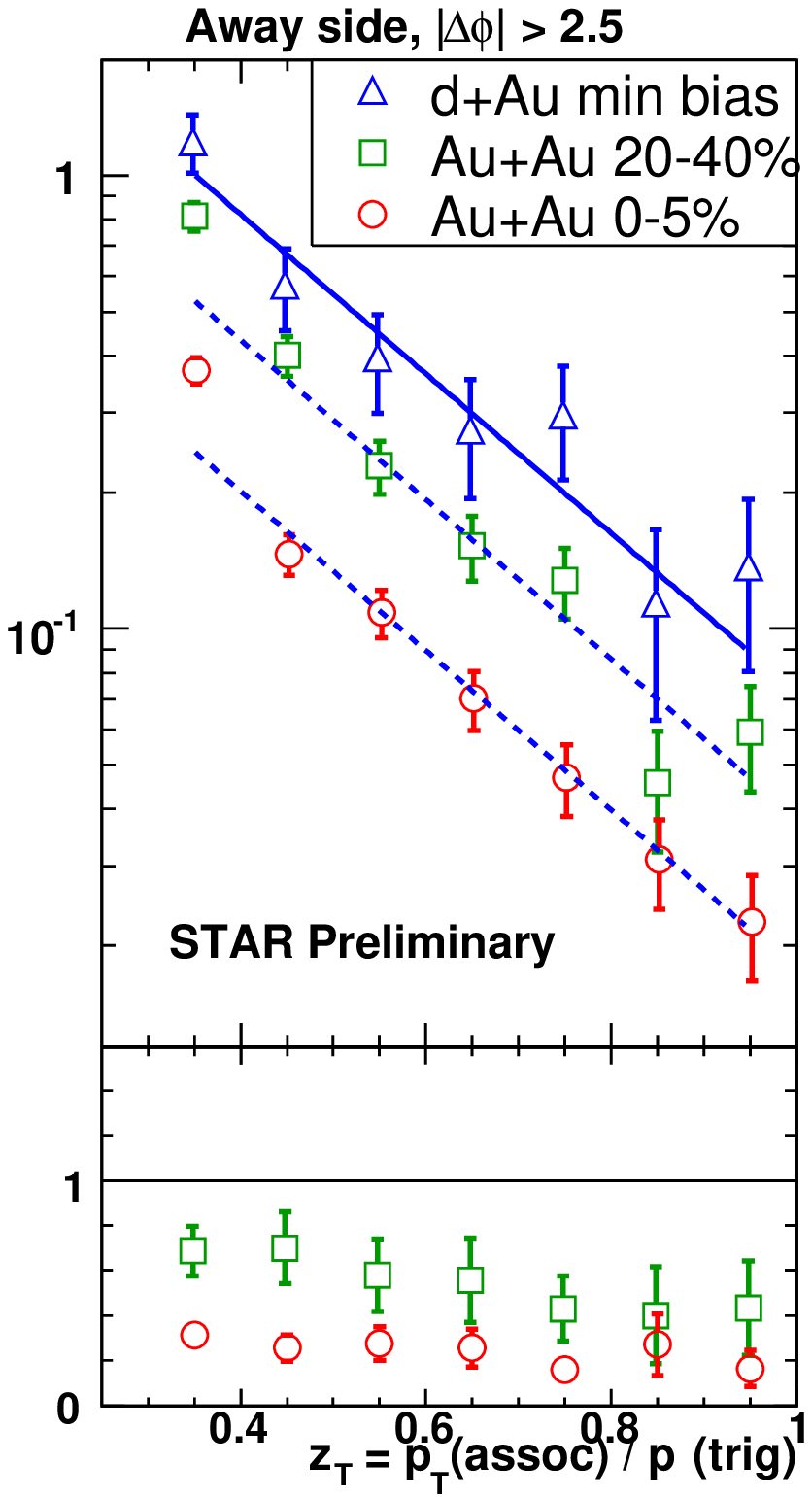}
  \includegraphics[height=.4\textheight]{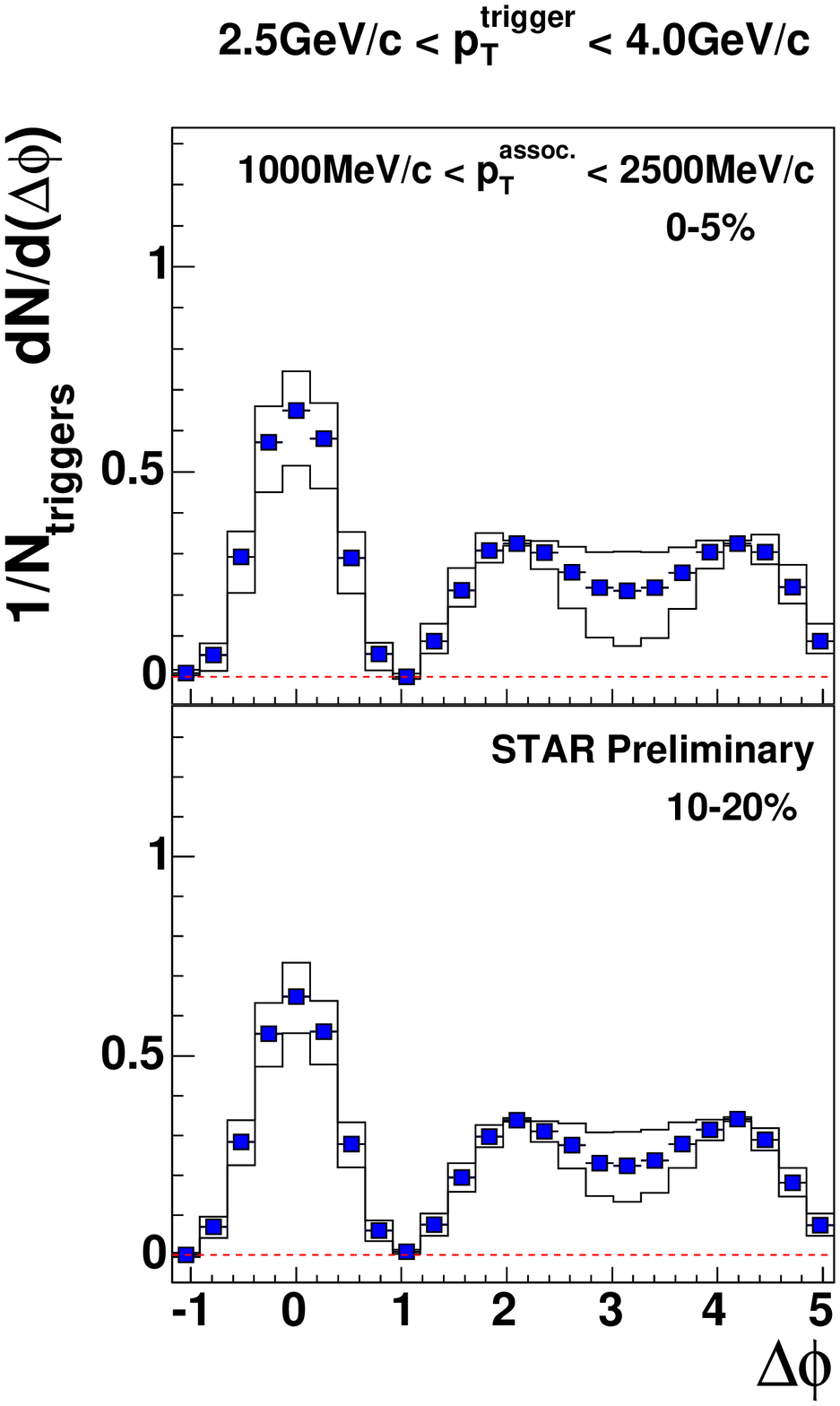}
  \caption{Left panel: Top: Hadron triggered
  fragmentation functions $dN/dz_T$ for away-side charged hadrons as a function of
  $z_T$ for 8 $< p_T^{trig} <$ 15 GeV/$c$ at $\sqrt{s_{_{NN}}}$ = 200 GeV $d$+Au and Au+Au
  collisions. The solid line is an exponential fit to the $z_T$ distribution for $d$+Au.
  The dashed lines represent the same exponential fit scaled down by factors of 0.54 and
  0.25 to approximate the yields in 20-40$\%$ and 0-5$\%$ Au+Au collisions. Bottom: Ratio
  of the hadron triggered fragmentation functions for
  Au+Au/$d$+Au. Right panel: Azimuthal distributions of associated
  charged hadrons for two different centralities at $\sqrt{s_{_{NN}}}$ = 200 GeV Au+Au
  collisions. The histograms indicate the systematic uncertainty
  bands.
  }\label{fig:azy}
\end{figure}

A well defined back-to-back peak that is characteristic of di-jets
is observed with negligible background in Fig. \ref{fig:jet} in
both peripheral and central Au+Au collisions measured by STAR
\cite{mag}. While the yield is substantially less in central Au+Au
collisions, the widths of the back-to-back peaks appear to be
independent of centrality. The away-side hadron triggered
fragmentation functions \cite{wan} for $d$+Au and Au+Au collisions
\cite{mag} measured by STAR as a function of $z_T =
p_T^{assoc}/p_T^{trig}$ is depicted in Fig. \ref{fig:azy} (left
panel). Fig. \ref{fig:azy} (left panel) shows that the shape of
the away-side fragmentation functions is unchanged from $d$+Au to
central Au+Au collisions. However, the yields are reduced by a
factor of $\sim$4 in central Au+Au collisions. Even though the
shape is consistent with previous predictions in the $z_T$ range
measured, the magnitude is smaller than expected \cite{wan}. A
different calculation predicted that significant energy loss
should be associated with significant broadening of the away-side
hadron azimuthal distribution \cite{vit1}, in contradiction to the
STAR measurements.

When low-$p_T$ associated particles are observed in central Au+Au
collisions opposite a high-$p_T$ trigger particle, they are found
to be significantly broadened in $\Delta\phi$ and enhanced in
number compared to $p+p$ collisions \cite{prl95}. Contrary to the
maximum found in $p+p$ and $d$+Au collisions, STAR measured the
$\langle p_T \rangle$ of the associated particles appears to be a
minimum at $\Delta\phi = \pi$ in central Au+Au collisions
\cite{ule}. In addition, the away-side associated particle yield
is flat or may have a small dip at $\Delta\phi = \pi$ \cite{ule}.
These phenomena are displayed in Fig. \ref{fig:azy} (right panel)
and have led to predictions that we may be observing jets that
have been deflected by radial flow or a Mach cone effect
associated with conical shock waves \cite{sto}.

\section{Heavy Flavor}

\begin{figure}
  \includegraphics[height=.3\textheight]{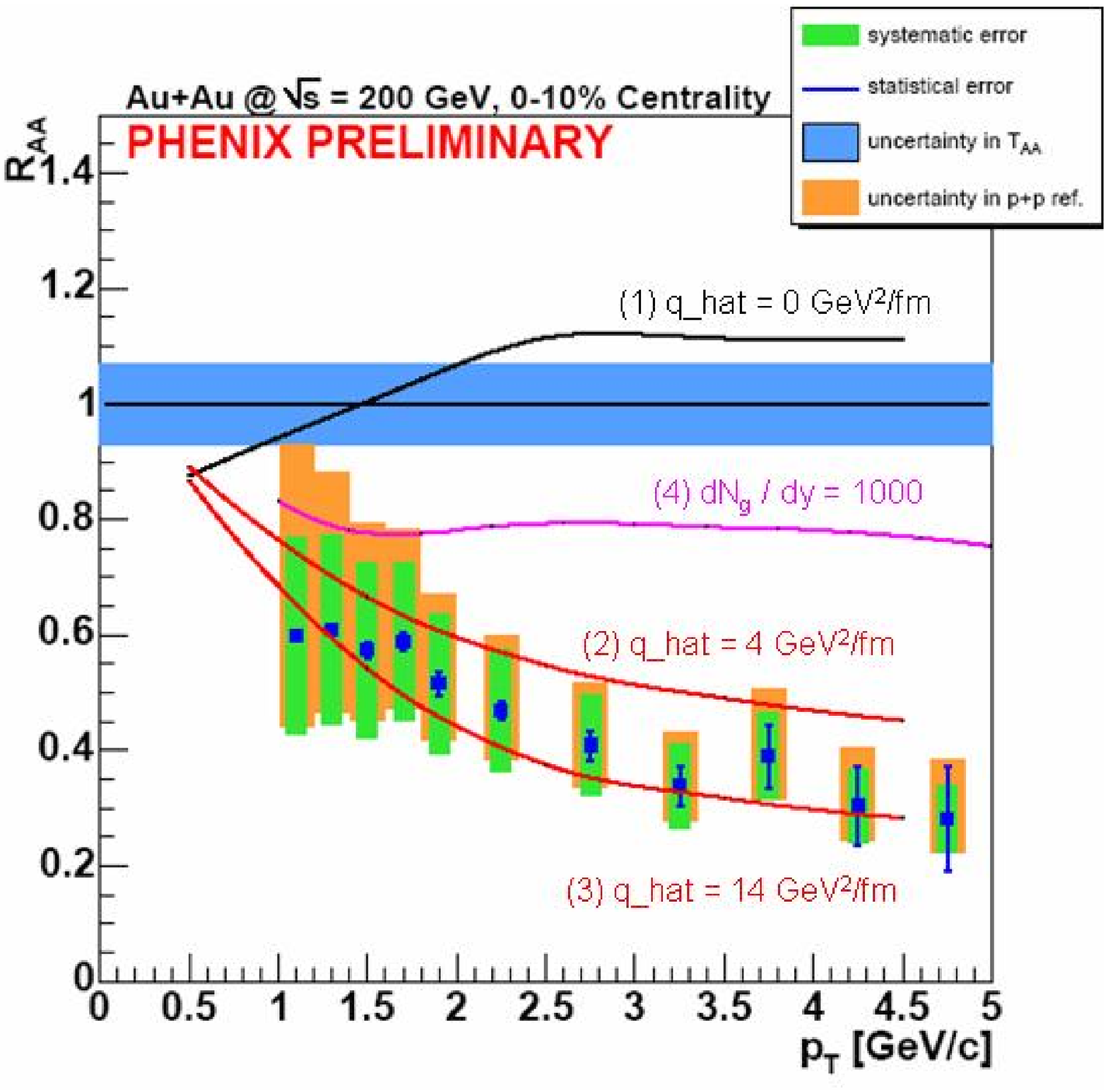}
  \includegraphics[height=.3\textheight]{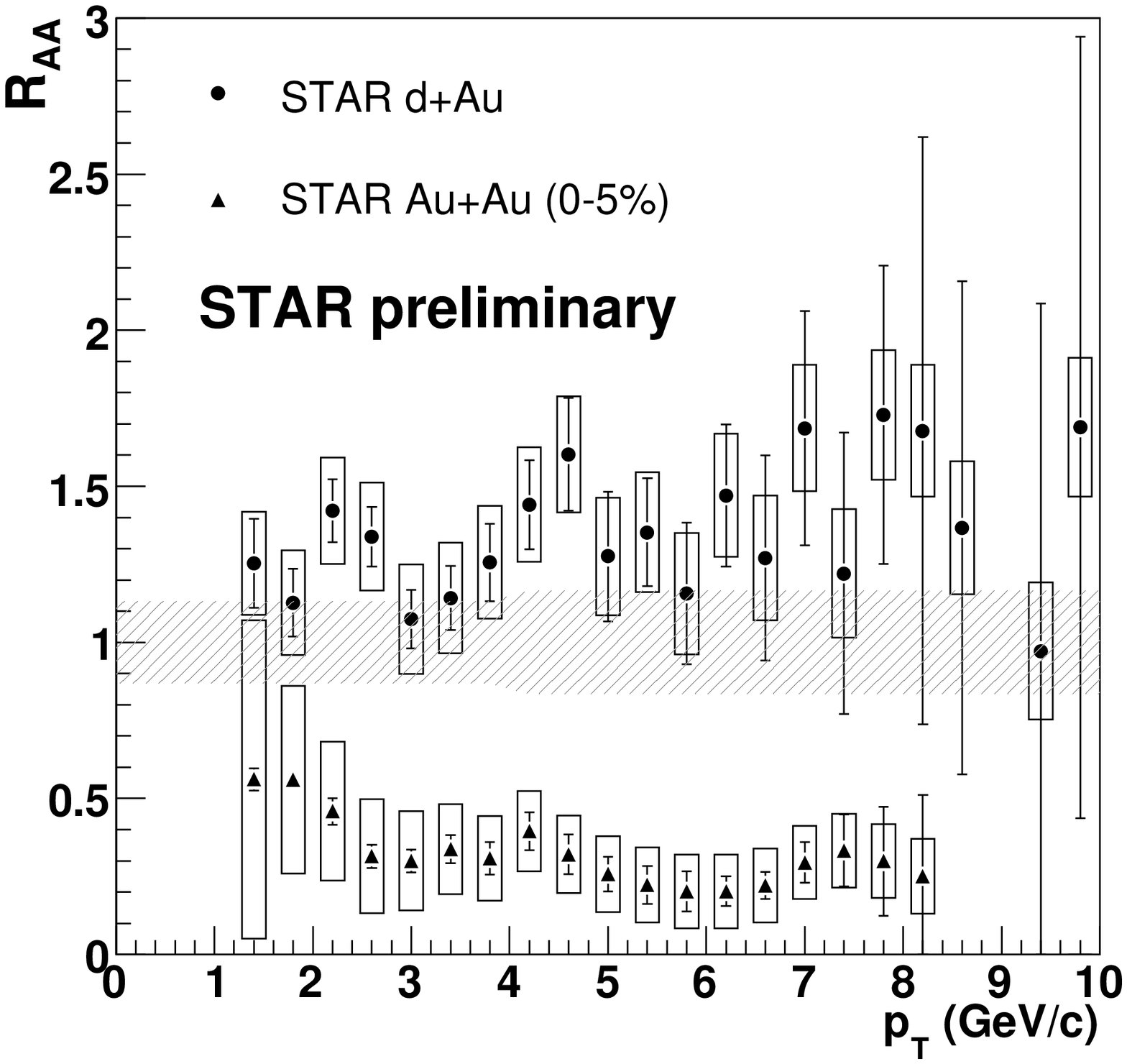}
  \caption{Left panel: $R_{AA}$ of non-photonic electrons measured by
  PHENIX \cite{but}
  for central Au+Au collisions at $\sqrt{s_{_{NN}}}$ = 200 GeV. Right panel:
  $R_{AA}$ of non-photonic electrons measured by STAR \cite{dun}
  for central Au+Au collisions at $\sqrt{s_{_{NN}}}$ = 200 GeV.}\label{fig:electrons}
\end{figure}

The nuclear modification factor $R_{AA}$ of non-photonic electrons
measured by PHENIX \cite{but} and STAR \cite{dun} in central Au+Au
collisions is shown in Fig. \ref{fig:electrons}. Both experiments
observe a very strong suppression of the non-photonic electrons.
The suppression has approximately the same shape and magnitude as
the suppression for hadrons, which is a quite surprising result,
since the massive quarks are expected to radiate much less energy
than the lighter $u$ and $d$ quarks. It is important to confirm
and complement these results with direct measurements of open
charm and open bottom contributions. The first direct
reconstruction of $D$ mesons by STAR \cite{zha} is the first step
in this direction, and more improvement will come with the
implementation of new detectors from both PHENIX and STAR.

\begin{figure}
  \includegraphics[height=.29\textheight]{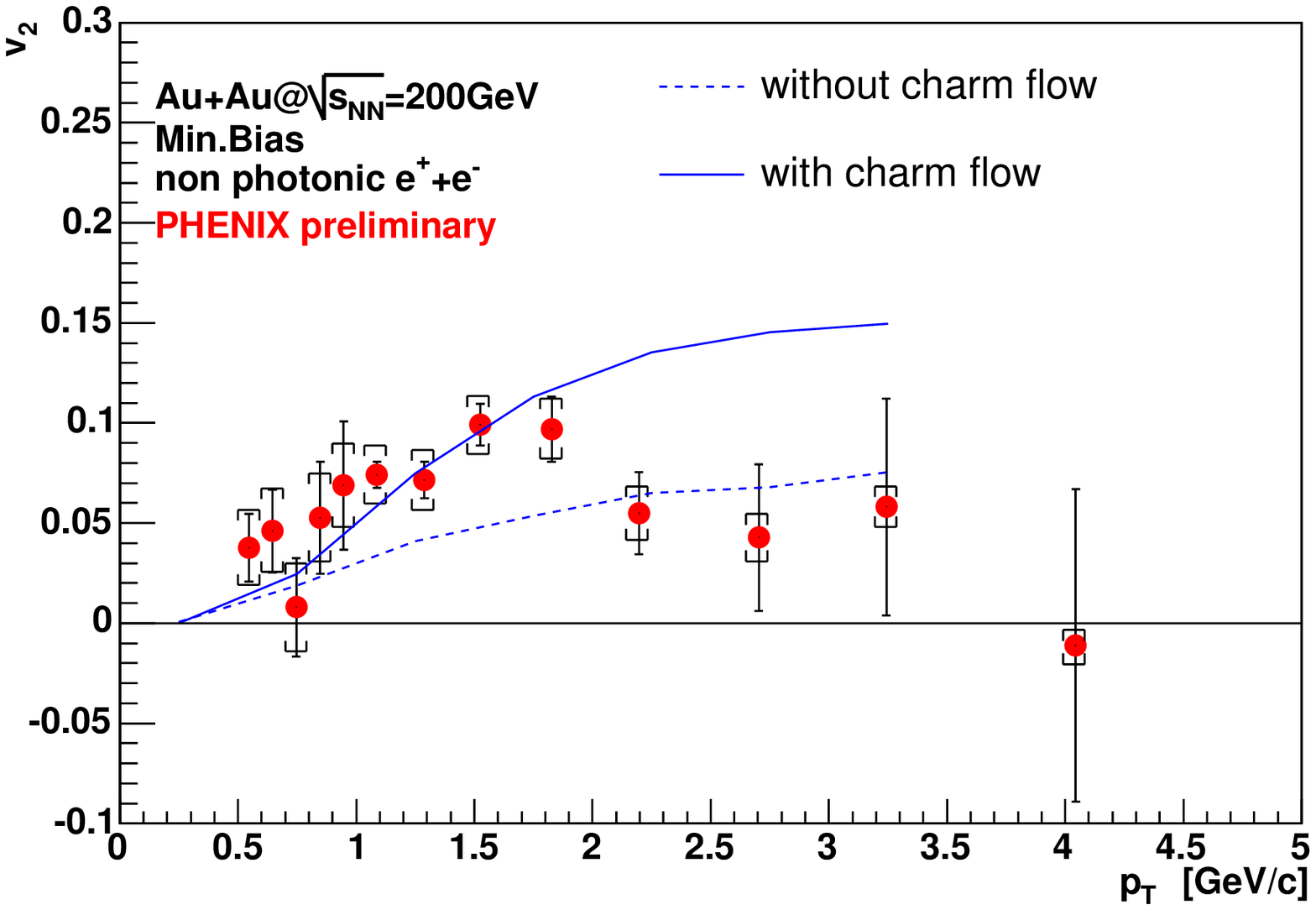}
  \includegraphics[height=.36\textheight]{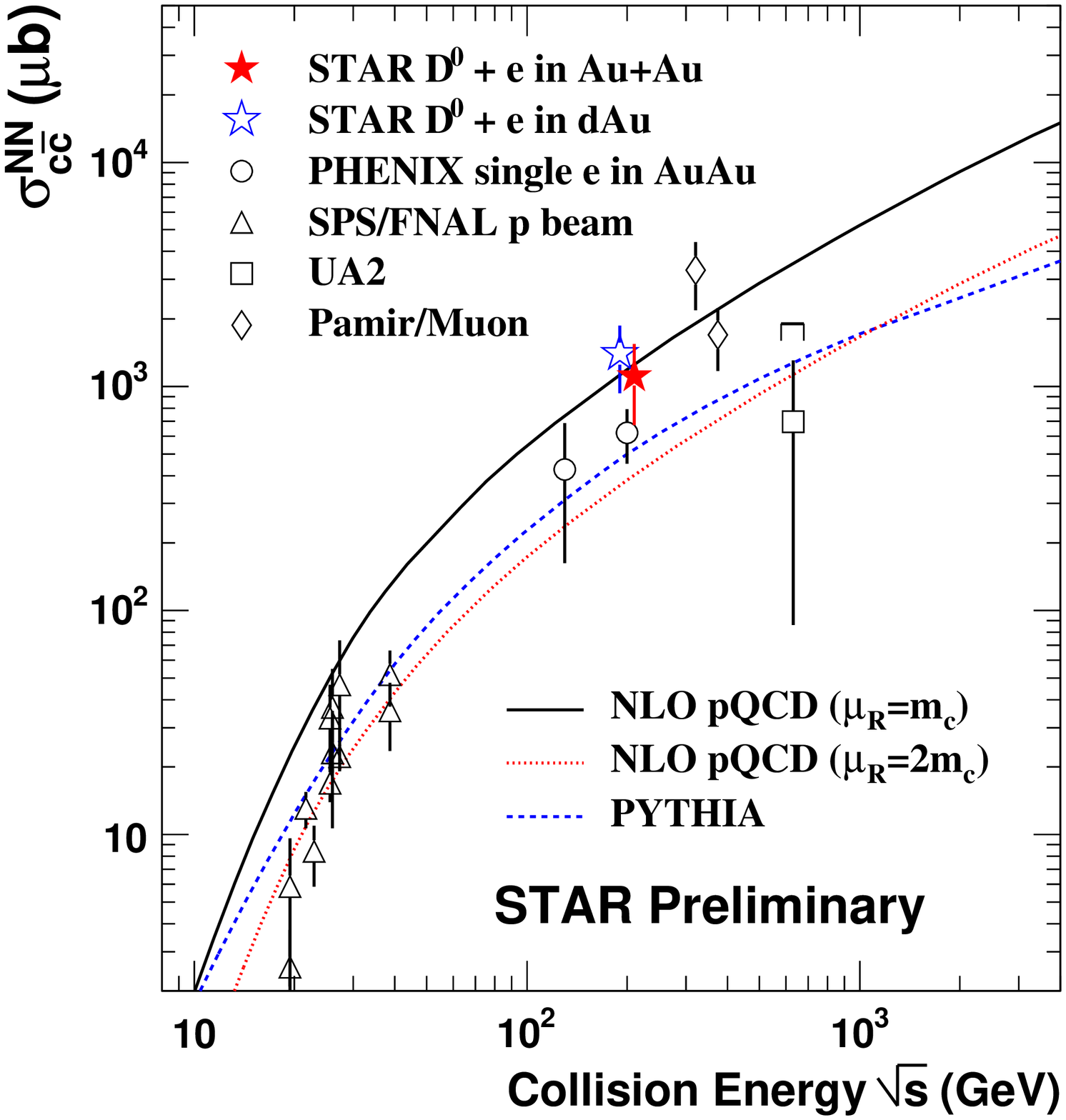}
  \caption{Left panel: $v_2$ of non-photonic electrons attributted to semi-leptonic
  open charm decays measured by PHENIX \cite{but} and compared to theoretical predictions
  \cite{gre1}. Right panel: Compilation of the total charm cross-section
  production per nucleon collision $\sigma_{c\bar{c}}^{NN}$
  as a function of energy compared to the NLO pQCD calculations \cite{zha}.}\label{fig:charm}
\end{figure}

The $v_2$ of non-photonic electrons as a function of $p_T$
\cite{but} measured by PHENIX is shown in Fig. \ref{fig:charm}
(left panel). The non-zero flow measured at $p_T <$ 2 GeV/$c$,
where the yield is dominated by semi-leptonic decays of open
charm, suggests a sizeable flow of $D$ mesons. The comparison to
calculations with and without charm flow \cite{gre1} shown in Fig.
\ref{fig:charm} (left panel) favors the interpretation of charm
flow. The consequence is a strong interaction with the medium and
a high degree of thermalization of $c$ quarks, favoring the
strongly coupled QGP.

PHENIX has calculated the non-photonic electron measurements in
$p+p$ and minimum Au+Au collisions at $\sqrt{s_{_{NN}}}$ = 200 GeV
into a total cross-section of charm production per nucleon
collision $\sigma_{c\bar{c}}^{NN}$ \cite{adl}. STAR has calculated
the combined electron and direct $D$ meson measurements in minimum
bias $d$+Au and minimum bias Au+Au collisions \cite{zha}. There is
a factor of $\sim$2 difference in their minimum bias Au+Au
cross-sections. It is important to notice that STAR is sensitive
to $80\%$ of the total charm cross-section, while PHENIX is only
sensitive to $15\%$. A compilation of charm cross-sections
$\sigma_{c\bar{c}}^{NN}$ as a function of collision energy is
displayed in Fig. \ref{fig:charm} (right panel). Results from
PHENIX and STAR are also shown.

\section{J/$\psi$}

\begin{figure}
  \includegraphics[height=.33\textheight]{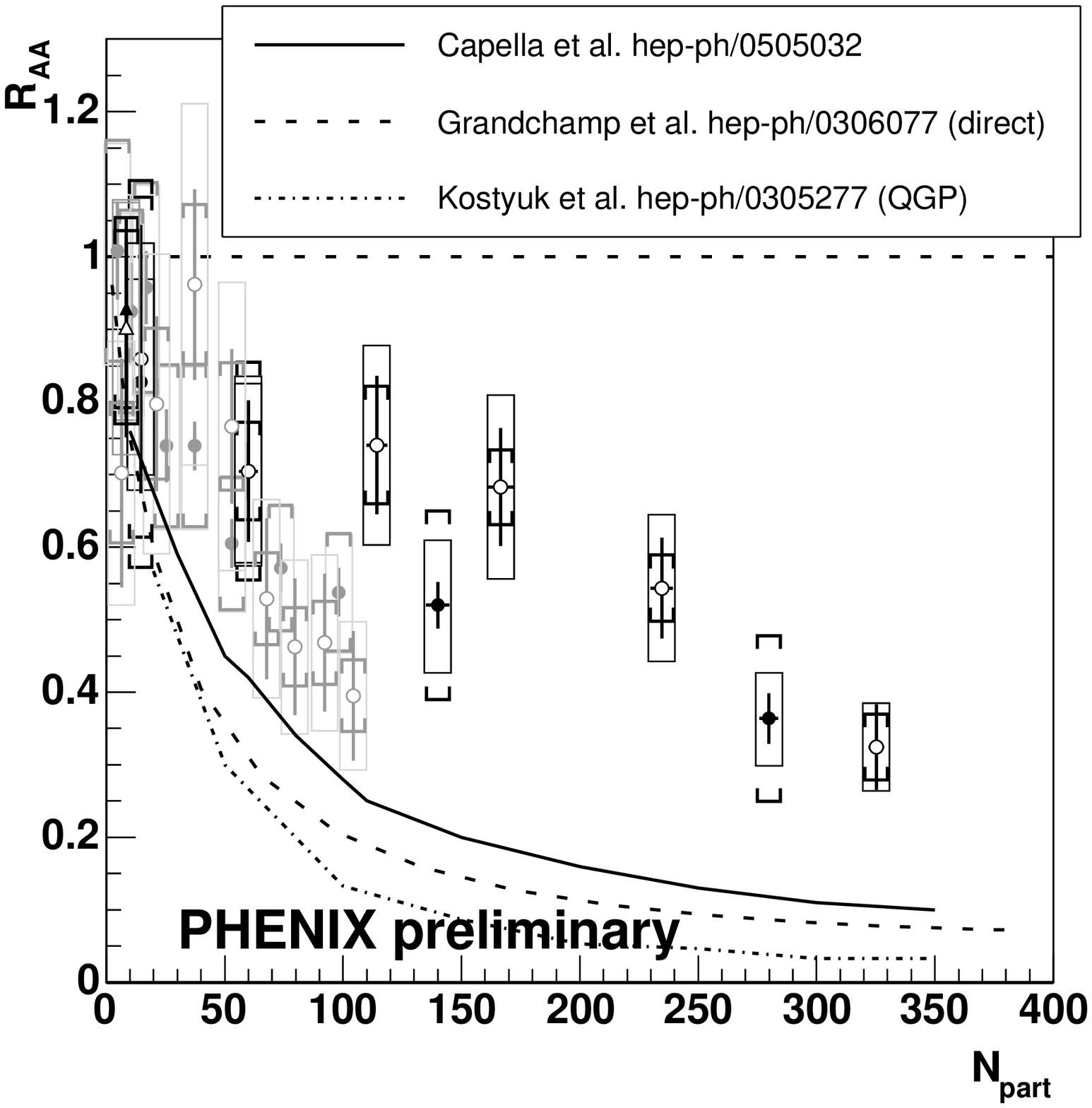}
  \includegraphics[height=.33\textheight]{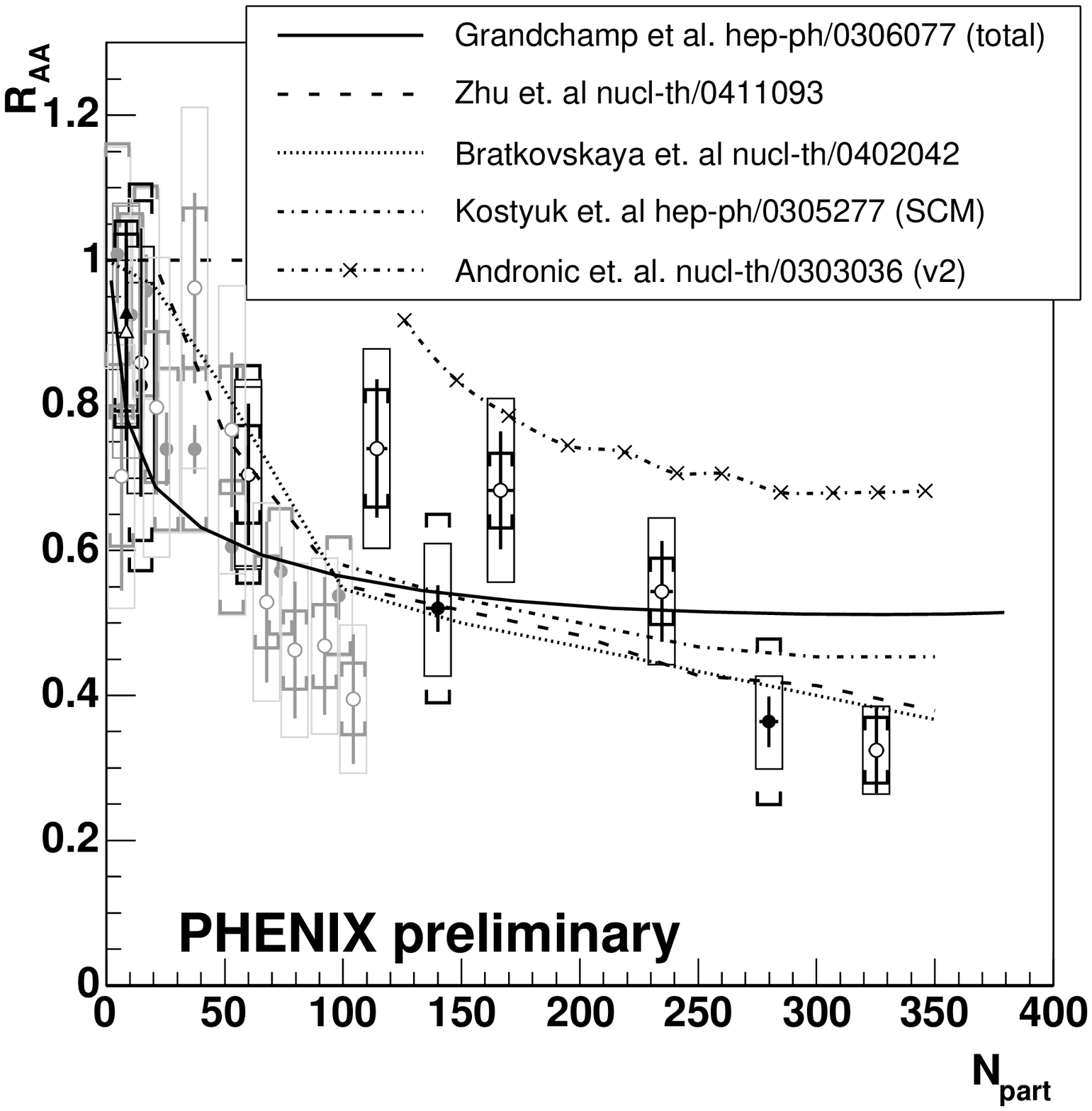}
  \caption{$R_{AA}$ of $J/\psi$ as a function of $N_{part}$ in
  $d$+Au, Cu+Cu, and Au+Au collisions at $\sqrt{s_{_{NN}}}$ =
  200 GeV measured by PHENIX. Left panel: The measurements are
  compared to models that explain the $J/\psi$ NA50 anomalous
  suppression \cite{cap,kos,gra}. Right panel: The measurements are
  compared to models involving either $J/\psi$ regeneration by quark
  recombination \cite{kos,gra,bra,and} or $J/\psi$ transport in medium
  \cite{zhu}.}\label{fig:jpsi}
\end{figure}

PHENIX measures $J/\psi$ in $p+p$, $d$+Au, Cu+Cu, and Au+Au
collisions at $\sqrt{s_{_{NN}}}$ = 200 GeV and Cu+Cu at
$\sqrt{s_{_{NN}}}$ = 62 GeV at mid-pseudorapidity $|\eta| < 0.35$
through the $e^+e^-$ decay channel  and forward pseudorapidity
$|\eta| \in [1.2,2.2]$ through the $\mu^+\mu^-$ decay channel
\cite{bue}. The $R_{AA}$ as a function of centrality $N_{part}$
for the measurements at $\sqrt{s_{_{NN}}}$ = 200 GeV are depicted
in Fig. \ref{fig:jpsi}. The suppression is clear and within errors
it is seems to be  independent of the collision system.
Furthermore, the magnitude of the suppression of $\sim$3 for the
most central collisions is similar to the suppression observed at
SPS \cite{tse}.

The left panel of Fig. \ref{fig:jpsi} shows that models that were
able to explain the anomalous $J/\psi$ suppression at the SPS and
that were based on interactions with comovers \cite{cap}, color
screening \cite{kos}, or QCD-inspired in-medium effects \cite{gra}
predict a stronger suppression at RHIC.

The discrepancy seems to be resolved by invoking the regeneration
of $J/\psi$ at the later stage of the collision via recombination
of $c$ and $\bar{c}$ quarks, which are produced more abundantly at
RHIC. Several attempts that combine suppression and recombination
\cite{kos,gra,bra,and} reproduce the data reasonably well, as
shown in the right panel of Fig. \ref{fig:jpsi}.

\section{Low-Mass Di-electron}

\begin{figure}
  \includegraphics[height=.29\textheight]{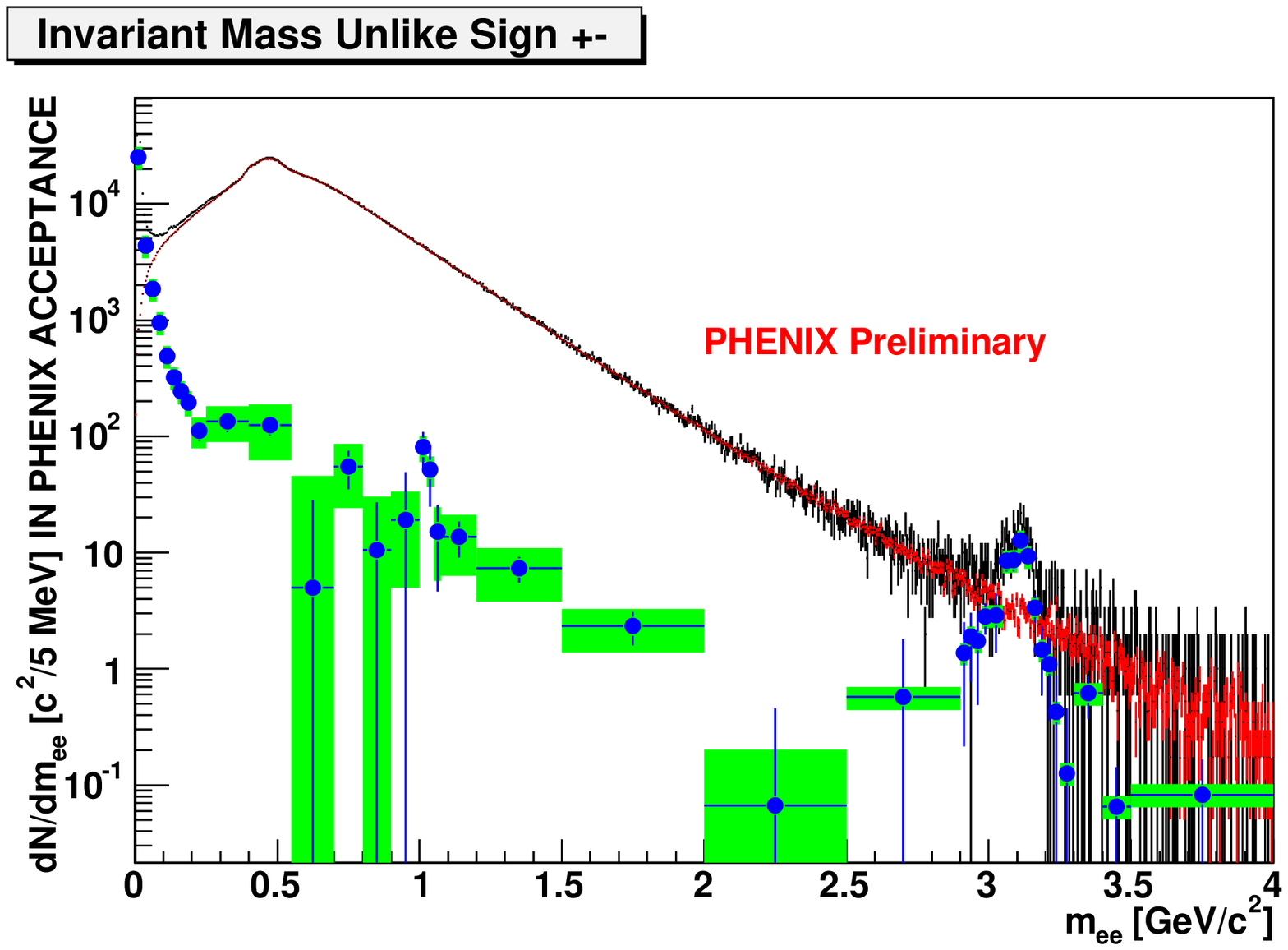}
  \includegraphics[height=.36\textheight]{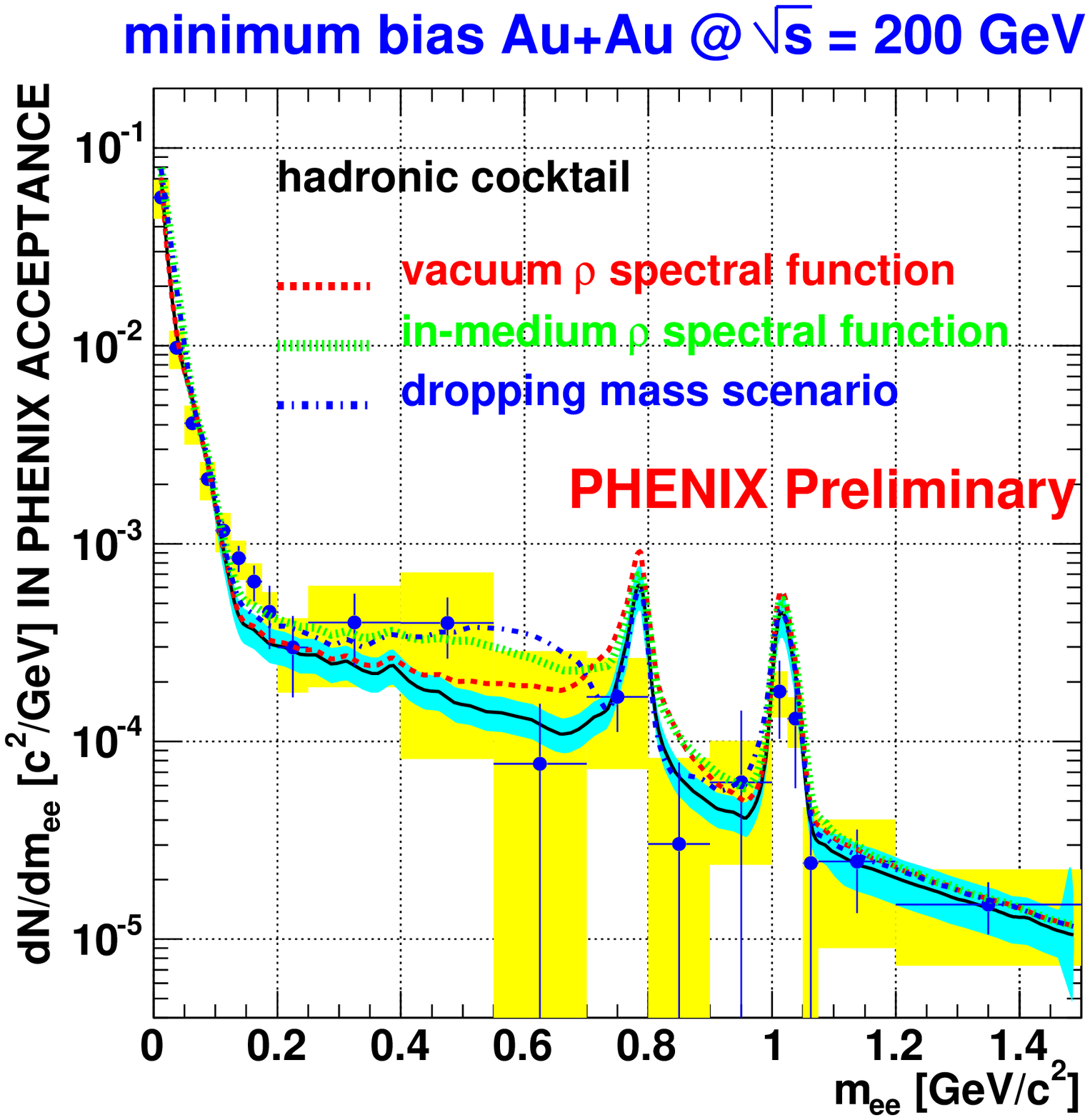}
  \caption{Left panel: The unlike sign mass spectrum together with
  the subtracted spectrum. Right panel: The data compared to the hadronic
  cocktail \cite{toi}, and to theoretical predictions
 \cite{rap}, where a $\rho$ spectral
 function is introduced with and without in-medium
 modifications.}\label{fig:dileptons}
\end{figure}

The left panel from Fig. \ref{fig:dileptons} depicts the measured
 di-electrons pairs, the background, and the subtracted spectra with
 uncertainty from PHENIX \cite{toi}. The right panel of Fig.
 \ref{fig:dileptons} shows the data compared to the hadronic
 cocktail \cite{toi}, and to theoretical predictions
 \cite{rap}, where the $e^+e^-$ invariant mass spectra have
 been calculated using different in-medium $\rho$ spectral
 functions and an expanding thermal fireball model. It is still
 premature for any statement, but it is definitely promising.

\section{Summary}

The baryon-meson effect observed at RHIC favors the constituent
quark coalescence as the production mechanism at intermediate
$p_T$, suggesting a partonic state prior to hadronization. $\phi,
\Xi$, and $\Omega$ have small hadronic cross-sections, therefore
the large $v_2$ observed suggest that it is built up in the
partonic stage. Furthermore, the thermal parameters extracted from
these particles suggest that they chemically and kinetically
freeze-out at the same time.

Intermediate $p_T$ ($p_T <$ 6 GeV/$c$) protons behave differently
than mesons at heavy-ion collisions. The large enhancement of the
$p/\pi$ ratio at intermediate $p_T$ in Au+Au collisions indicates
that jet fragmentation in vacuum is not the dominant source of
particle production in this $p_T$ range.

The suppression of high $p_T$ hadrons in central $Au+Au$
collisions was one of the unexpected and important phenomena
observed at RHIC. The suppression is quite strong and remains
approximately flat up to 20 GeV/$c$. Partonic radiative energy
loss models \cite{vit,wan,esk} reproduce this behavior well.

The quality of the RHIC data is just marvellous, and the knowledge
that this data has provided on the dense matter formed in
relativistic heavy-ion collisions is without any doubt unexpected.
We can only wonder how new data with new detector upgrades will
surprise us.

\begin{theacknowledgments}
I would like to thank F. Laue, R. Longacre, T. Ullrich, Z. Xu, and
H. Zhang for the exciting discussions. This work was supported in
part by the HENP Divisions of the Office of Science of the U.S.
DOE.
\end{theacknowledgments}

\end{document}